\documentstyle[aps,prl,twocolumn]{revtex}
\begin{document}
\twocolumn[\hsize\textwidth\columnwidth\hsize\csname
  @twocolumnfalse\endcsname
\preprint{IMSc 98/04/13}
%\preprint{hep-ph@ftp/9804204}
\title{Perturbative QCD with string tension}
\author{Ramesh Anishetty \cite{email}}
\address{The Institute of Mathematical Sciences, Chennai - 600 113, INDIA}
\maketitle
\begin{abstract}
String tension is introduced in  perturbative QCD without hindering
ultra violet renormalizability . The running gauge  coupling constant remains
unaltered and the string tension in the high energy limit  vanishes
asymptotically. Using 1/N techinique   the theory can be explicitly  analytically
computable with cofinement built in . Spontaneous chiral symmetry
breakdown along with its Goldstone is exhibited.
\end{abstract}
\vskip 2pc] % end \twocolumn[...]
%\pacs{11.10St,11.15Tk,11.15Pq,11.30Rd,12.38Ge,12.38Lg}
We  present a simple picture of confinement with in perturbation
theory which can readily make quantitative predictions in the framework of
renormalizable relativistic quantum field theory. It was perhaps Feynman who
first suggested that  color confinement may come about because of a singular
$1\over{q^4}$  gluon propagator where $q$ is the momentum of the gluon. Many attempts 
to deduce this in QCD have  given several scenarios such as dual superconducting
vacuum \cite{mand1}, 
singular gluon propagator as a possible solution to the Schwinger Dyson
equation \cite{baker}, lattice QCD \cite{wils}. We will
draw our attention to lattice pure  non-abelian gauge theories. These theories
are the simplest to understand in the strong coupling limit where they exhibit
linear confinement. The continuum theory that these describe is in the weak
coupling regime. Numerical simulations  \cite{creu} have given the picture ,using
renormalization group techiniques, that in the weak coupling they are indeed
describing the continuum perturbative QCD quantitatively, namely the much
celebrated running coupling constant. In the very same regime lattice gauge
theory has non vanishing string tension, the numerical dimensionful parameter
which is the coefficient of linear potential between heavy quark and antiquark.
The lattice theory is therfore describing  continuum theory in the weak
coupling regime with string tension. We ask ourselves the question can we
directly describe this effective field theory in the continuum  language.

In this paper we would like to demonstarte the existence of effective theories
in continuum which have string tension with in perturbation theory. It turns
out these theories are also renormalizable, hence quantitative predictions can
be made. For example the running gauge coulping constant is exactly the same as
in the standard QCD. The string tension also runs in such a way that it vanishes
at very high momenta. There are many other which can be looked into, here we
will make a preliminary attempt to work out some of the important consequences
in the presence of light quarks. Using 1/N techinique  \cite{thooft1} we exhibit spontaneous
chiral symmetry breakdown and the Goldstone thereof.

Consider pure QCD without quarks. The generating functional for
arbitrary source current $j_ \mu ^a$ is given in terms of the following
Euclidean functional integral
\begin{equation}
\int DA_\mu ^a exp{(-\int d^4x({ 1\over {4g^2}} {F_{\mu \nu} ^a }^2 +
 i j_ \mu ^a A_ \mu ^a)}
\end{equation}
where $ F_{\mu \nu} ^a = \partial _\mu A_\nu ^a - \partial _\nu A_\mu ^a
  +f ^{a b c } A _\mu ^b A_\nu ^c$
and $g$ is the gauge coupling constant and $f^{abc}$ are the structure
constants of the corresponding non-abelian gauge group. To this action defined in terms of the
gauge fields $A$  alone,  we cannot add any other
term if we want to preserve renormalizability. In particular there is no room
to add a dimensionful parameter namely the string tension $\sigma$ .
To motivate the inclusion of $\sigma$ we enlarge our field configuration space
by making the following formal rewriting
\begin{eqnarray}
exp{(-i \int j.A)} = \int DQ \delta (Q-A)exp{(-i \int j.Q)} 
 =\\
\int DQ DC D \overline \chi D \chi exp{-i \int
(CD^2(Q-A)+\overline \chi D^2 \chi + j.Q)}
\end{eqnarray}
where $Q$ and $C$  are bosonic vector fields and $\overline{\chi}$ and $\chi$ are
Grassmann Lorentz vector fields. All of them are in adjoint representation of
the gauge group and transform covariantly under local gauge transformations.
$D_\mu ^{a b} = \partial _ \mu \delta ^ {a b} + f ^ {a c b } A _ \mu ^c $
is the gauge covariant derivative operator.
(2) follows from (3) by noting that -$D^2$ is a positive operator, therefore
integrating over the Lagrange multiplier field $C$ yields the constraint $Q=A$. The
$\chi$ fields integration will compensate the functional measure to yield (2).

Formally using (3) functional integral (1) is re-expressed in
a larger field configuration space. All the newly introduced fields  by
construction are of 
dimension one by tree level power counting. Without losing renormalizability
we introduce string tension namely by adding a term $\sigma C^2$ to the 
Lagrangian. Shifting $Q$ to $Q+A$ we collect all the terms and define a new
generating functional
      
\begin{eqnarray}
Z = \int DA DQ DC D \overline \chi D \chi exp{( -S_0 -i \int j.( Q+A) )}, 
 \nonumber \\  S_0 = \int ({ 1 \over 4g^2} F ^2 +i C_ \mu D^2 Q_ \mu
 +i \overline \chi _\mu D^2 \chi _\mu + {\sigma \over 2} C_ \mu ^2 )
\label{eq:S0}
\end{eqnarray}
A remark in order, 
the $\sigma$ term in (\ref{eq:S0}) can be interpreted as being generated due to quantum
fluctuations in standard QCD.  However, it can be seen that the $\sigma$ term
cannot be generated in the small $g$ 
perturbation theory. This is because in the absence of $\sigma$ term there are
two global supersymmetries( not space-time) of the action in (\ref{eq:S0}) following
from the invariance  of $S_ 0$under the variations:
\begin{eqnarray}
\delta _ \theta \chi _\mu = \theta Q _\mu  \quad ; \quad 
 \delta _\theta C _\mu = \theta\overline \chi _ \mu \\ 
 \delta _{\overline \theta} \overline \chi = \overline \theta C _\mu \quad 
 ; \quad \delta _{\overline \theta} Q _\mu = \chi _\mu \overline \theta 
\end{eqnarray}
where $\theta$ and $\overline{\theta}$ are two arbitrary Grassmann parameters.
In perturbation theory these supersymmtries cannot be spontaneously broken.
Hence in QCD the sigma term can only be generated by non-perturbative
mechanisms for example in dual superconducting vacuum, in which case the $\sigma$
parameter will be functionally related to the gauge coupling constant $g$ and
the QCD scale parameter.

Here we can  take the view that (\ref{eq:S0}) is an effective field theory for QCD 
alternatively (\ref{eq:S0})
  is a fundamental renormalizable theory in its own right. Much of what
we are going to discuss can entertain both view points. There are many more 
renormalizable terms that can be added to (4). Most of these
are dimension four terms and are quite harmless but will make algebra more
cumbersome. There are two dimension two terms which we will address shortly.

First we demonstrate linear confinement in Gaussian perturbation theory for (\ref{eq:S0})
.We need to do gauge fixing and the associated ghost fields: this aspect is
completely identical to the conventional perturbative QCD. The new propagators
in momentum space are

\begin{eqnarray}
< C _\mu ^a Q _\nu ^b > = {i \eta _ {\mu \nu} \delta ^{a b} \over {q^2}}
  = < \overline \chi _\mu ^a \chi _\nu ^b > \\
 < Q _\mu ^a Q _\nu ^b > = \sigma { \eta _{\mu \nu} \delta ^{a b} \over {(q
^2)^2}} \quad ; \quad  <C_ \mu ^a C_\nu ^b > = 0  \label{eq:QQ} 
\end{eqnarray}
The external source interacts with both $Q$ and $A$ vector fields in the form
prescribed by
(\ref{eq:S0}) . The $<QQ>$ propagator having $1/q^4$ behaviour will give linear confinement
to any external source even in small $g$ perturbation theory. Using dimensional
regularization we can do perturbation theory. In addition to the standard gluon
interactions we have gluon interactions with $Q$ and $C$ vector fields and also
with $\overline{\chi},\chi$ fields.

First look at any 1PI  graph with
external $A$ legs: In addition to the standard QCD graphs we will also have other
graphs wherein $<CQ>$ propagator loop can also come in . To any order all
these loops cancel exactly by construction with $\overline{\chi},\chi$ loops.
Note that we cannot have any $<QQ>$ propagtor in these loops  since the basic
interactions are either $ACQ$ or $AACQ$ and all $C$ and $Q$ vertices can only be saturated
with $<CQ>$ propagators because  $<CC>$ vanishes (8).Therefore to all loops all
1PI graphs with external $A$ legs do not have any modification finte or
infinite(i.e. ultra violet regularization dependant) from that of standard
QCD.Consequently the "$\beta$ function", $\beta(g)$ remains unchanged 
and is also independent of $\sigma$. 
 
Next look at connected 1PI graphs with $Q$ and $A$ as external legs such as $Q^2$ , $Q^4$,
$(DQ)^2$ etc. These are also manifestly absent by noting that $<CC>$ vanishes
(\ref{eq:QQ}).
These consequences can easily be understood by noting that first $\sigma$
term breaks supersymmetry (5) softly but $\overline \theta $ supersymmetry
(6) remains  unbroken. The latter is sufficient to enforce the above 
conclusions. This is an important consequence for if $Q^2$ term can be induced
at any loop level the string tension interpretation of $\sigma$ ceases.

Higher order loops do indeed generate term such as $D_\mu C_\mu D_\nu Q_\nu +
D_\mu {\overline \chi}_\mu D_\nu \chi$ which are supersymmetric , these do
not make any qualitative difference to our conclusions although for
computational brevity we  will choose a renormalization prescription that they
vanish. Similarily we will renormalize terms like $(D_\mu C_\mu)^2$ to zero.

Then there remains the term $ C.Q + {\overline \chi } . \chi$ which is
supersymmetric. At one loop level, in dimensional regularization it happens to
be absent. At higher loops this can come around. Presence of
this term destroys confinement picture. However we do have the choice of
renormalizing this term to zero as well. This should be interpreted in the
following manner. In the field configuration space as defined by the fuctional
integral (\ref{eq:S0}), of all the possible action functionals we are exploring the regime which exhibits
linear confinement , to achieve this goal we make the suitable renormalization.
 
   Now let us look at the renormalized $\sigma$ in the ultra violet region.
The calculations are very simple to do in the background gauge. In this gauge
we need to make wave function renormalization for $C$ ,$\overline \chi$ and $A$ fields.
Doing so using dimensional regularization we obtain the running string
tension $\sigma (\mu)$ at the scale $\mu$ is given by the following equation
\begin{equation}
{\partial \sigma (\mu) \over {\partial ln \mu ^2}} = 
-6\sigma {\alpha _s 
\over {4 \pi}} C_2  \quad ; \quad
 {{\partial {1/ {\alpha _s} }\over 
{ \partial ln \mu ^2}}}
 ={11 C_2 \over {3 (4 \pi)}} 
\end{equation} 
where $C _2$ is the quadratic Casimir in the adjoint representation of the
gauge group and $\alpha _s = g^2 /4 \pi $ is the running coupling constant.
Solving the above ,we obtain the running string tension in terms of the standard
running gauge coupling $\alpha_s (\mu)$,
\begin{equation}
\sigma (\mu)= \sigma (\mu _0 ) [{\alpha _s (\mu ) \over 
{\alpha _s (\mu
_0) }} ] ^{18/11} 
\end{equation}
(10) shows that at  high energies quantum fluctuations make the string tension 
to vanish asymptotically  at a rate faster than the running coupling constant.

Now we  couple the theory defined by (\ref{eq:S0})  to dynamical quarks. Minimal
gauge principle is now modified uniformly according to the source functional
(\ref{eq:S0}) 
by replacing all covariant derivatives by $D(A+Q)$. Therefore  the total
action $S$ for chiral
symmetric quark interaction with gluons is given by 
\begin{equation}
S= S_0 +i\int {\overline q} \gamma_
\mu D_ \mu  (A+Q) \quad q  \label{eq:S}
\end{equation}
Quark fields $q$ and $\overline q$ are in the fundamental representation hence
the corresponding covariant derivative is implied in (11).
This interaction breaks both supersymmetries (5) and (6). Consequently loop
corrections can induce $Q^2$ terms, this in turn reflects that the propagator
 $<QQ>$ becomes that of a massive particle,
that is linear confinement is no longer present. This is just the
phenomenon that we are familiar in lattice gauge theories, namely in the presence
of dynamical quarks, the string breaks into quark and anti-quark pairs and hence
the linear confinement ceases.

Here we make a remark that to one loop we explicitly find that no $Q^2$
counterterm is  generated.  Similarily the $\beta$ function also does not get any
additioanl contributions up to two loops. We suspect that to all orders  the
$\beta$ function remains unchanged from that of standard QCD. The argument being $\sigma$ is a dimension
two parameter consequently its ultra violet contribution is suppressed by two
powers of the ultra violet regularizing scale which most certainly cannot
modify the the logarithmic scale  that the $\beta $ function is capturing. 

Now we would like to reinterpret our action $S$ in terms of the physcially
interesting fields alone. 
 The total action including quarks is linear in the $Q$ fields. So this can be
explicitly integrated to yield a functional constraint  which in turn can be
solved for $C$ fields. We also integrate the remaining redundant fields
$\overline \chi$ and $\chi$ , then we obtain a quartic non-local Nambu Jona Lasinio typ
e of interaction \cite{namb},
\begin{equation}
{\sigma \over 2}
\int {\overline {q} \gamma _ \mu T^a q}[{1\over {(D^2)^2 }}]^{ab}{\overline {q}
 \gamma _ \mu T^b q} \label{eq:nj}
\end{equation} 
where $\overline {q} \gamma_\mu T^a q$ is the local color current due to quarks
which is interacting non-locally with itself. This interaction is manifestly
invariant under local gauge transformations and global chiral rotations.

Therfore our theory can be reinterpreted in terms of basic quark and gluon
fields as a theory given by the standard QCD action 
modified by  the above non local interaction which manifestly has confinement built in.
It is surprising to see that such a non local theory can be analyzed with
standard perturbative renormalization techiniques by recasting into the form(11)
that we have considered in this paper.(\ref{eq:nj}) gives us the interpretation that
$C,Q,\overline {\chi}, \chi$  are all  auxillary fields which allow
us to define the formal functional  integral in perturbation theory.

We now  study the quark sector of the theory
given by $S$. For example
whether the quark propagator exhibits spontaneous symmetry breakdown. This
question, in principle can be answered by solving the Schwinger Dyson equation
, but in practice the equation involves higher order vertices on which we have
less control.
 't Hooft \cite{thooft1} ,\cite{thooft2} showed that a pragmatic approximation which
is consistent with phenomenonlogy and make systematic approximation to QCD
 is to consider the number of colours N to be an another
expansion parameter. We find that in our context 1/N expansion makes many of
our computations practically feasible. This expansion is defined by the
following limits $g^2,\sigma \rightarrow 0$ and N $\rightarrow \infty$ such
that  $g^2$N and $\sigma$N are kept fixed.

The theory (\ref{eq:S}) has three expansion parameters $\sigma$N , $g^2$N and 1/N, of 
these the last
is necessarily small. In addition we consider the case when $g^2$N is also small.
Hence the leading term in our expansion is non-perturbative in the parameter
$\sigma $N. In this limit also we would be summing an infinite number of
Feynman graphs which are essential for binding of colored fields to singlets.
In this limit all interactions are mediated by $<QQ>$ propagator (\ref{eq:QQ})
 .$Q$ is a
covariant non-abelian field but  has no self-interactions in our
approximation .

Consider the quark propagtor in our systematic approximation scheme.
The Schwinger Dyson equation sums the so called "rainbow graphs " with vector
boson propagator taken to be $<QQ>$. The full quark propagator $S(p)$ satisfies
the integral equation
\begin{equation}
S^{-1} (p)= \gamma .p + \sigma N \int {d^4k \over {(2\pi)^4}}{1 \over {(k^2)^2}}
\gamma _ \mu S(p-k) \gamma _\mu  \label{eq:sd}
\end{equation}
Make the following spinor decomposition $S(p)=\gamma  .pA(p^2 )+B(p^2)$ we
can re-express (\ref{eq:sd}) as two coupled equations for the two scalar functions $A$ and $B$.
If $B$ is non-zero it signifies that our vacuum has spontaneously broken chiral
symmetry.  Now look into  how  the Goldstone is realized in our
approximation.
The Bethe-Salpeter equation for
pseudoscalar $\phi _5$,
($\overline {q} \gamma_ 5 q $),only ladder graphs involving $<QQ>$ and the 
full quark propagator $S(p)$  contribute, given by
\begin{eqnarray}
S^{-1}(p-q)\phi _ 5 (p,p-q)S^{-1}(p) =\sigma N \int {d^4 k \over {(2 \pi)^4}} \nonumber
\\ {1\over {(k^2)^2}}(\gamma _ \mu \phi _ 5 (k+p,k+p-q) \gamma _ \mu 
 +...)\label{eq:bs}
\end{eqnarray} 
where $q$ is the total momentum of the bound state, $p$ and $p-q$ are the
momenta of quark and anti-quark.
Additional contributions,in (14) which vanish for small $q$
are  represented by dots. We expect an eigenfunction corresponding to
$q=0$. Indeed from (13)  by formal algebra we verify that $\phi _5 (p,p) = 
\gamma _5 B(p^2) $ is a solution to (\ref{eq:bs}). This property of the Bethe Salpeter
and Schwinger Dyson equations is true in very general terms for spontaneously
broken theories, in the context of
"rainbow graphs"  approximation it was first realized by Mandelstam
\cite{mand2}.This
exercise demonstrates that the 1/N approximation defined by taking $g^2N$ small
and $\sigma N$ arbitrary is a consistent approximation to study chiral
symmetry. Furthermore we find that to leading order the spin independent quark
propagator is identical to the pion wave function , a result which is perhaps
testable.

(\ref{eq:sd}),( \ref{eq:bs}) are coupled non-linear integral equations. These we are able to
solve analytically as a series expansion in the corresponding momenta. These results will be discussed in a future publication.
Here we remark on an important techinical issue. The singular $<QQ>$
propagator introduces logarithmic infra red divergences, these need to be
regularized. Regularization can be achieved formally by a simple procedure:
Introduce a $\delta$ term as $\delta (C.Q +\overline {\chi}. \chi )$. into our action
(11)
then all singular propagators (7) and (8) become those of a massive one. Then all the infra
-red logs are regularized. Then we can adopt the standard renormalization
procedure to remove these infra-red logs and in the end the renormalized
$\delta$ is taken to be zero.
 
We quote one important quantity 
\begin{equation}
S(0)=B(0)= {{2 \pi} \over \sqrt{\sigma N}}
\end{equation}
as a solution from (\ref{eq:sd}) showing chiral symmetry breakdown.

Explicit chiral symmetry breaking term  i.e. mass term for the quarks can be studied . We have looked into
various other vector mesons. Formal analytical solution  for boundstate 
wave functions can be deduced. All
these will be addressed in a separate paper. Baryons need some extra care and
this is being presently persued. In our approximation various scattering
processes can be considered. With the gauge coupling assumed to be small we can
explicitly consider various scattering contributions. The picture that emerges
is very much like 't Hooft model \cite{thooft2}, \cite{coleman} in the leading order  
where the confining string tension is taken into consideration
non-perturbatively and to this there are
additional contributions  of the order $g^2N$ coming from gluons, $<AA>$ propagators just like in
parton model with a  qualitative difference i.e.  our calculations make sense
even for soft momenta because of explicit meson wave functions that there are
in our leading approximation.

Finally we remark about the overall picture. Standard QCD may have confinement
only in the strong  coupling limit which is also a very non trivial theory.
But the most important quantity which captures the qualitative behaviour is the
string tension. Now we may suspect that in a theory with the same symmetries as
standard QCD but with an explicit string tension at the tree level can capture
the strong coupling standard QCD with  an
effective gauge coupling which is weak. Such a scenario is
envisaged and an explicit proposal is being made in this paper. Some
preliminary attempt to understand the phenomenonlogy  such as the $\beta $
function, chiral symmetry breakdown is found to be
encouraging.     

 In the theory defined by (11) within the 1/N and $g^2$N small approximation
we have no glueballs. This is because $A$ fields never interact through
singular $<QQ>$ propagator. Consequently no gluons can be binded sufficiently
strongly. In our approximation explicit weak coupling massless gluons are still
present. These aspects need to be looked into more critically.

\end{document}